\documentclass[12pt]{article}

\usepackage{scicite}
\usepackage{times}

\usepackage{graphicx}
\usepackage{dcolumn}
\usepackage{bm}
\usepackage{siunitx}[separate-uncertainty, multi-part-units=single] 
\usepackage[version=3]{mhchem}
\usepackage{xcolor}

\newenvironment{sciabstract}{%
\begin{quote} \bf}
{\end{quote}}

\title{Fast optoelectronic charge state conversion of silicon vacancies in diamond}

\author
{Manuel Rieger,$^{1}$ Viviana Villafañe,$^{1,2}$ Lina M. Todenhagen,$^{1}$ Stephan Matthies,$^{2}$ \\ Stefan Appel,$^{2}$ Martin S. Brandt,$^{1}$ Kai M{\" u}ller,$^{2}$ Jonathan J. Finley$^{1\ast}$\\
\\
\normalsize{$^{1}$Walter Schottky Institute, School of Natural Sciences and MCQST,}\\ 
\normalsize{Technical University of Munich, 85748 Garching, Germany}\\
\normalsize{$^{2}$Walter Schottky Institute, School of Computation, Information and Technology and MCQST,}\\ 
\normalsize{Technical University of Munich, 85748 Garching, Germany}\\
\\
\normalsize{$^\ast$ finley@wsi.tum.de}
}

\date{17.10.2023}

\begin{document} 


\baselineskip24pt


\maketitle


\begin{sciabstract}
    Group IV vacancy color centers in diamond are promising spin-photon interfaces with strong potential for applications in photonic quantum technologies. Reliable methods for controlling and stabilizing their charge state are urgently needed for scaling to multi-qubit devices. Here, we manipulate the charge state of silicon vacancy (SiV) ensembles by combining luminescence and photo-current spectroscopy. We controllably convert the charge state between the optically active SiV$^-$ and dark SiV$^{2-}$ with \SI{}{MHz} rates and $>\SI{90}{\%}$ contrast by judiciously choosing the local potential applied to in-plane surface electrodes and the laser excitation wavelength. We observe intense SiV$^-$ photoluminescence under hole-capture, measure the intrinsic conversion time from the dark SiV$^{2-}$ to the bright SiV$^-$ to be \SI{36.4\pm6.7}{\milli\second} and demonstrate how it can be enhanced by a factor of $10^5$ via optical pumping. Moreover, we obtain new information on the defects that contribute to photo-conductivity, indicating the presence of substitutional nitrogen and divacancies.
\end{sciabstract}


\section*{Introduction}

Color centers in diamond are a promising platform for quantum sensing and quantum network applications \cite{nguyen2019quantum, bhaskar2020experimental, pompili2021realization, PhysRevX.13.011042}. Group IV vacancy centers (G4V) in diamond, including silicon (SiV), germanium (GeV) and tin (SnV) vacancies, have exceptional optical properties such as a bright zero-phonon lines, large Debye-Waller factors and lifetime-limited linewidths \cite{Bradac2019_GV4review,Dietrich2014_SiV_sideband,Goerlitz2020_SnV_phononSideband,Zhou2017_SiV_lifetimeLimited}. Due to their inversion symmetry along the bond axis, G4V complexes are less susceptible to electric noise enabling their integration into diamond nanophotonic structures to create efficient spin-photon interfaces \cite{Sipahigil2014_Indistinguishability,GaliMaze,Sipahigil2016_integratedCavity}. Among the different G4V centers, the negatively charged silicon vacancy (SiV$^-$) is the most intensively studied center. It shows excellent coherence times in the range of \SI{10}{\milli\second} for electron spins and several hundred \SI{}{\milli\second} for nuclear spins \cite{sukachev2017silicon}. 
The interaction of G4V complexes with proximal impurities or defects in the diamond host can be detrimental for their charge and spectral stability, possibly hindering their use as a spin-photon interface \cite{TinVacancyCycle_Becher2022}.
Unlike nitrogen vacancy (NV) centers, SiV do not efficiently cycle between two charge states when pumping them with visible light. This indicates weaker interaction between the trapped electron and one of the continuum bands of the host diamond \cite{lozovoi2022imaging}.
The photo-physics of G4V-centers remains subject to intense debate focused around whether trapped electrons are photo-excited via the conduction or valence bands \cite{Nicolas2019_SiV-bandDiagram-bothHypothesis} and whether the state under charge conversion is the bright neutral \cite{Zhang2023_SiV0stabilization_photoitinerantCarriers,Dhomkar2018_SiV0-on-demand-doubtable} or dark doubly negatively charged state \cite{Gardill2021_probingSiV-withNV,TinVacancyCycle_Becher2022}. For SnV, the consensus has converged more towards interaction with the valence band under illumination in the visible \cite{TinVacancyCycle_Becher2022} and our results support this hypothesis also for SiV. Whilst SiV$^-$ is a bright emitter, SiV$^{2-}$ is optically dark and SiV$^0$ emits at a different frequency \cite{Wood2023_SiV0_SiV2-_Meriles}. The charge state depends on the nature and concentration of proximal defects, the illumination scheme and the proximity of electric contacts that influence carrier capture and ionization rates via the electric fields they generate.
Several strategies have previously been proposed to control the SiV charge state, including doping \cite{Rose2018_SiV0_250msCoherence}, chemical surface treatments \cite{Zhang2023PRL_SiV0-by-H-termination, grotz2012charge}, and integration into p-i-n diodes to control the position of the quasi Fermi level  \cite{Bray2020_PIN-Diode-SiV,Tegetmeyer2016_PINdiode_Electroluminescence}. 

In this paper, we present investigations of the opto-electronic interconversion between the optically dark SiV$^{2-}$ and bright SiV$^-$ charge state. For this, we define interdigitated metallic contacts on the diamond surface to facilitate the application of local electric fields and combine with quasi-resonant optical illumination to efficiently and reversibly control the SiV charge state.  Our results show that the charge state can be converted at fast (\SI{}{MHz}) rates under optical illumination by controlling the electrical bias and can be applied to a specific center located within a device, opening the way toward fully integrated geometries. 
We examine the conversion for synthetic, undoped, commercially implanted electronic grade single-crystal diamonds typically used by the quantum photonics community. Specifically, we find that the excitation wavelength providing the maximum ($>\SI{90}{\%}$) conversion contrast for the lowest charge conversion voltage to be \SI{2.4}{\eV}, readily accessible using diode-pumped solid state lasers and, moreover, show that this is related to a deeper resonance of the SiV$^-$ from the lower-lying $a_{2u}$ level \cite{Haußler2017_a2u_level,GaliMaze}. 
Using time-resolved photoluminescence measurements we demonstrate that the non-equilibrium SiV$^{2-}$ ensemble relaxes to SiV$^-$ with a characteristic exponential decay time of \SI{36\pm7}{\milli\second} under dark conditions, while it converts within $<$\SI{0.25}{\micro\second} under illumination, more than 5 orders of magnitude faster. We show that the $\mathrm{SiV}^{2-}\rightarrow\mathrm{SiV}^-$ conversion is facilitated by capturing optically created holes from the valence band. Maxima in the ensemble SiV$^-$ PL at the positively biased electrode indicate ensemble charge state stabilization. 

\section*{Results}

\subsection*{Photoluminescence at zero electrical bias and photocurrent signal}

\begin{figure*}
\centering
\includegraphics[width=1\textwidth]{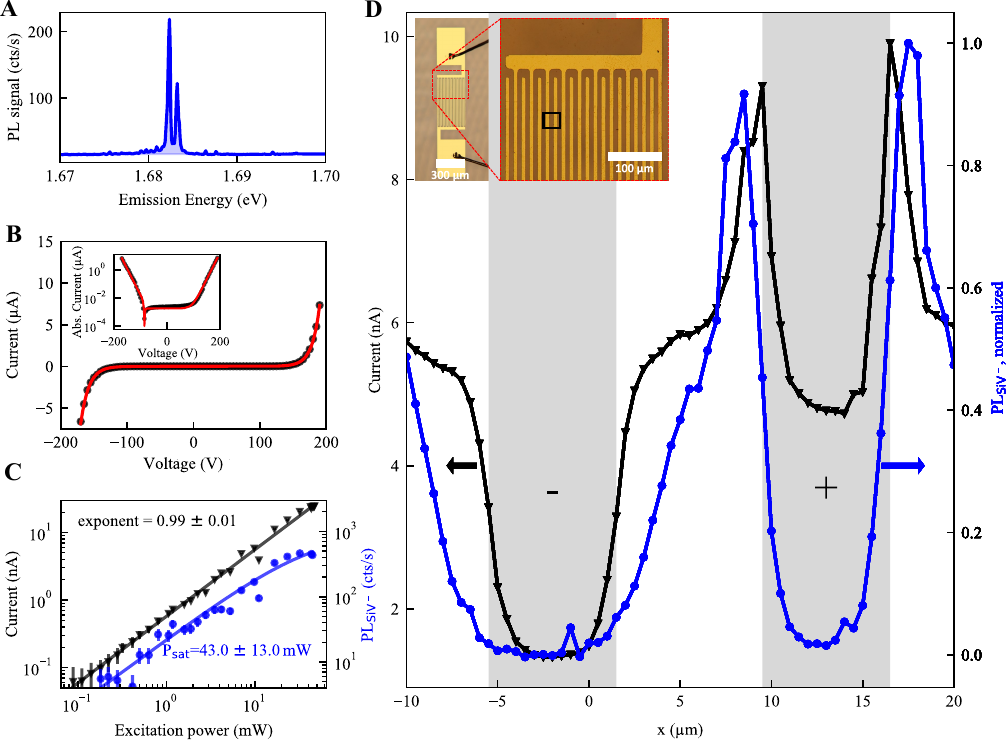}
\caption{\label{fig:SiV-PL_sample_Map-cut}
\textbf{Spatially-resolved SiV$^-$ photolumincescence and photocurrent.}
\textbf{A} Typical photoluminescence (PL) spectrum of the SiV$^-$ ensemble obtained under non-resonant excitation with a \SI{2.40}{\eV} continuous-wave (CW) laser. The characteristic \SI{250}{\giga\hertz} excited state splitting of the SiV$^-$ is clearly visible. The grey shaded area marks the integrated SiV$^-$ photoluminescence intensity, PL$_{\rm{SiV^-}}$ plotted in other panels of the figure.
\textbf{B} Photocurrent versus bias voltage when excited with \SI{1.84}{\eV} pulsed laser light. The inset shows the absolute current in log scale. The voltage dependent current is fitted using equation \ref{eq:PC}.
\textbf{C} The current (triangles) and the SiV$^-$ PL intensity (circles) scale approximately linearly with the green CW excitation laser power $P_{\mathrm{ex}}$, while the photoluminescence intensity shows saturation at higher $P_{\mathrm{ex}}$. We used \SI{+100}{\volt} DC bias for the PC measurement and illuminated the gold-diamond interface.
\textbf{D} Spatially-resolved dependence of the SiV$^-$ PL (circles) and  PC (triangles) under an electrical bias of \SI{160}{\volt} between the electrodes.
The grey shaded areas indicate the position of the (in-plane) electrodes and their polarity is marked by the plus and minus sign.
Both the current and the SiV$^-$ PL peak at the interface of the positive electrode with the diamond and decrease towards the negative electrode. We can therefore convert the SiV$^-$ PL by changing the bias polarity, which indicates a conversion to another SiV charge state.
}
\end{figure*}

As described in detail in the materials and methods section, the sample investigated is a high-purity, CVD grown single-crystal diamond that was implanted with $^{28}$Si to produce SiV centers at a depth from $100$ to \SI{150}{\nano\meter} below the surface. After annealing using a three step high temperature process under high vacuum \cite{Evans2016_annealingProcedure}, the sample was cleaned and equipped with an array of \SI{7}{\micro\meter} wide interdigitated Ti/Au contacts with an electrode-electrode spacing of \SI{8}{\micro\meter} (see the Materials and methods section for details).

Figure \ref{fig:SiV-PL_sample_Map-cut}A shows the characteristic photoluminescence (PL) spectrum of SiV$^-$ at $\sim$\SI{1.68}{\eV} \cite{Neu2013_lowTempPLSiV,Hepp2014_PL_SiV,Evans2016_annealingProcedure} recorded at \SI{10}{\kelvin} when subject to off-resonant excitation using a continuous wave green laser at photon energy of \SI{2.33}{\eV} (areal power density of \SI{180}{\watt/\centi\meter^2}). We do not observe the zero-phonon line of negatively charged nitrogen vacancy (NV$^-$) centers at \SI{1.95}{\eV} and only a very weak line at \SI{2.16}{\eV} (see figures S2 and S3 in the Supplemental) corresponding to an amount of neutral NV centers negligible in comparison to the number of SiV centers. The presence of NV$^0$ and absence of NV$^-$ in comparison with the small concentration of Boron indicates that there is another electron acceptor present in the system, which receives electrons from substitutional nitrogen. SiV centers are such an acceptor, with energy levels close to the valence band (VB) \cite{GaliMaze}. We also do not observe any indication of the neutral silicon vacancy (SiV$^0$) emission at \SI{1.31}{\eV}\cite{DHaenens-Johansson,Rose2018_SiV0_250msCoherence,Zhang2023PRL_SiV0-by-H-termination}, both under green and under pulsed near IR excitation at \SI{1.55}{\eV}. The fact that silicon vacancies occur mainly in the negative and possibly in the doubly negative (dark) charge state is typical for high-purity samples since neutral silicon vacancies require significant boron (or other p-type) doping \cite{Rose2018_SiV0_250msCoherence,Flatae2020}. Indeed, according to density functional theory (DFT) calculations, the presence of SiV$^-$ indicates weak p-doping. We note that most defects in diamond are deep in the band gap such that thermalization at cryogenic temperatures occurs only on very long timescales, so that photoexcited carriers dominate the charge state conversion processes \cite{Dhomkar2016_NV-long-time-chargeState-stability}.

We begin by discussing photocurrent (PC) measurements.  Hereby, we focus the excitation laser over several electrodes (\SI{28}{\micro\meter} $1/e^2$ beam waist, centered between two electrodes) to generate a PC signal that dominates over the dark current even at moderate excitation laser power $P_{\mathrm{ex}}$ of several hundred \SI{}{\micro\watt}. Figure \ref{fig:SiV-PL_sample_Map-cut}B presents a typical voltage dependent PC. We modelled the PC signal in our device building on a double Schottky barrier model \cite{grillo2021current} and thereby describe the photo-generated current $I$ at each electrode $j=1,2$ using
\begin{equation}
\label{eq:PC}
  I_j = SA^{*}T^{2}e^{-\phi_j/(k_B T+h\nu)},
\end{equation}
where $S$ is the junction area, $A^*$  is  the  Richardson  constant, $T$ is the temperature, $k_B$ is the Boltzmann constant, and $h\nu$ is the photon energy of the laser. In this expression $\phi$ is the effective modified Schottky barrier height,
\begin{equation}
\label{eq:schottkybarrier}
  \phi_j = \phi_0 \pm eV\left(1-\frac{1}{n_j}\right),
\end{equation}
where $\phi_0$  is the ideal Schottky barrier at zero bias, $V$ is the voltage drop at each of the diamond-metal junctions, and  $n_j$ is the ideality factor for each contact forming the double Schottky barrier. Figure \ref{fig:SiV-PL_sample_Map-cut}B compares typical measured  PC data obtained with excitation at \SI{1.84}{eV} (\SI{675}{nm}) with a fit provided by equation \ref{eq:PC}. Very good agreement is obtained between experiment and this phenomenological model. From the linear dependence of the PC on $P_{\mathrm{ex}}$ at the gold-diamond interface under green excitation presented in figure \ref{fig:SiV-PL_sample_Map-cut}C, we infer that the excitation from the VB to above the metal Fermi level is a single photon process (see figure S4 in the supplemental for a visualization). Arguments for why the interaction between the VB and the metal are the most likely source of this current are presented in the next paragraph. This means that the separation between the VB and $E_F$ at the positive electrode is $\le$ \SI{2.4}{\eV}, such that excitation into the CB would requires photons with an energy $\geq\SI{5.5}{\eV} - \SI{2.4}{\eV}=\SI{3.1}{\eV}$. Thereby, the large band gap of diamond hinders efficient creation of free electrons at the metal interface.

According to DFT calculations, photons with an energy  $\geq$\SI{3}{\eV} are required to excite an SiV$^-$ electron into the conduction band \cite{Thiering2018_MagnetoOpticalSpectra_G4V}. Therefore, direct excitation with green light of an electron from the excited state of SiV$^-$ into the conduction band is also incompatible with a linear power dependence. Excitation from the SiV$^-$ excited state to an electron donor level is unlikely since it requires a wavefunction overlap of the two centers and we use high-purity single-crystal diamond with impurity concentrations in the lower range of parts per billion (see the Materials and Methods Section). Therefore, we conclude that the measured PC signal does not stem from a multi-photon electron-hole pairs creation mediated by the excited state of SiV. In contrast, we identify the linear response as arising from free holes generated in the diamond VB by exciting electrons from the VB into defect levels via single photon absorption.
Up to photon energies of $E_g/2=\SI{2.75}{\eV}$, it is not possible to excite a defect both from the VB and to the CB with a single photon process for each step. A cascade of a one-photon process and a two-photon process could exhibit a linear laser-power dependence, but that would require the one-photon process to be significantly slower, which is very unlikely. In summary, all the data suggests that the photocurrent does not stem from optically created electron-hole pairs, neither mediated by the SiV nor other crystal defect states.
The arguments against optical electron-hole-pair creation also hold for the PC in between the electrodes. Instead, we propose that the PC stems from free holes created by optical excitation of a VB electron into a defect level which is in accord with recent findings on SnV charge state stabilization \cite{TinVacancyCycle_Becher2022}.

We continue by investigating the dependence of photocurrent and photoluminescence on the position of the excitation laser relative to the electrodes, which reveals central characteristics of the SiV charge state conversion. All following measurements are performed with a 0.5 NA objective that focuses the laser on the sample surface with a $1/e^2$ beam waist of \SI{2.2}{\micro\meter}. In figure \ref{fig:SiV-PL_sample_Map-cut}D, as the excitation laser spot is spatially swept from the positive interdigitated electrode to the negative one, with a constant bias of \SI{160}{\volt}, we observe a monotonous decrease of SiV$^-$ PL and PC. Interestingly, the PC maximum is centered exactly at the diamond/electrode interface. In other words, charges are most efficiently created and captured in the direct vicinity of the \textit{positive} electrode. Electrons from the VB are energetically photo-excited above the Fermi level of the positive electrode, leaving holes in the VB, which drift towards the negative electrode. This process can be mediated by interface states \cite{Lozovoi2020PRL} or by defects in direct vicinity of the metal-diamond contact where electrons can directly transfer into the positive electrode. This picture is fully consistent with the higher PC amplitude when the laser is focused directly on the positively biased electrode (see figure \ref{fig:SiV-PL_sample_Map-cut}D).  We note that one might intuitively expect zero PC at this position due to low laser transmission through the metal. We calculate the transmission through the \SI{80}{\nano\meter} thick electrode to be $\eta\approx\SI{2}{\%}$ at \SI{2.4}{\eV} \cite{GoldTransmission}). 
The gold-diamond interface region immediately next to the electrode produces the highest PC signal, indicating that the large interface below the electrode also contributes significantly to the current, even though the laser is strongly attenuated below the gold layer. In contrast, when detecting PL both the excitation and the emitted light must pass through the electrode, and therefore, the PL intensity recorded from the electrode is expected to be $\propto\eta^2$, fully consistent with the data presented in figure \ref{fig:SiV-PL_sample_Map-cut}D.

\subsection*{Spectral dependence of the charge state conversion}

\begin{figure*}
\centering
\includegraphics[width=\textwidth]{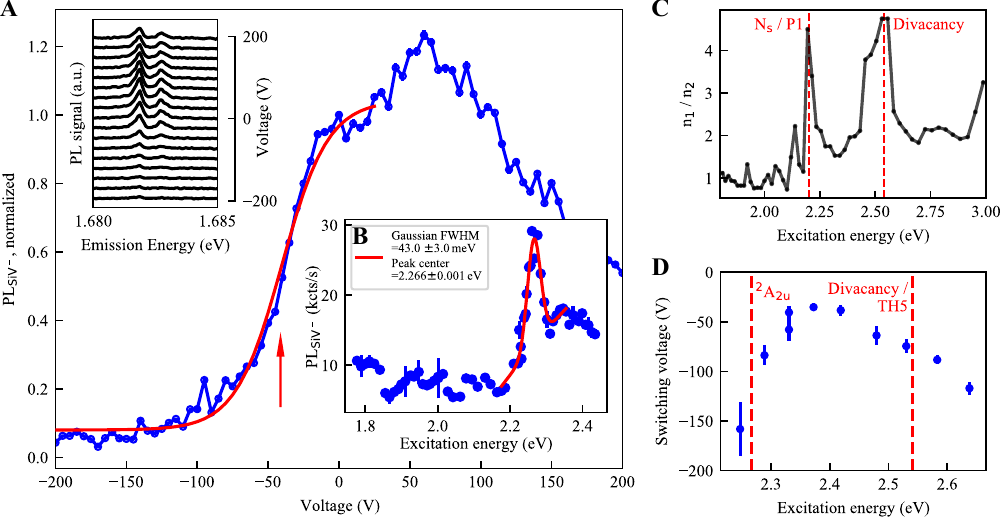}
\caption{\label{fig:wavelength-dependence-switching}
\textbf{Excitation energy dependence of the charge state conversion and defect spectroscopy.}
\textbf{A} Integrated SiV$^-$ PL versus bias voltage at the diamond-metal interface. For negative bias, the SiV$^-$ PL and hence fraction of total SiV can be described by a sigmoidal fit, where we define the center (marked with the arrow) as the conversion voltage. The PL intensity increases for positive bias up to \SI{50}{\volt}, which indicates stabilization/maximization of the negative (SiV$^-$) ensemble.
The inset on the top left shows some of the voltage dependent PL spectra that were integrated (after background subtraction) for the main plot.
\textbf{B} Integrated ensemble SiV$^-$ PL intensity as a function of excitation photon energy while exciting centered between two electrodes under moderate \SI{20}{\volt} bias. From a Gaussian fit with linear background we find a resonance for SiV$^-$ excitation at \SI{2.27}{\eV} and a full width at half maximum of \SI{43(3)}{\milli\eV}, which corresponds to resonant excitation to the $^2$A$_{\rm{2u}}$ higher excited state.
\textbf{C} Ratio of the fitted ideality factors extracted from photocurrent versus bias measurements under different excitation wavelengths and a \SI{10}{\micro\meter} laser spot size. The ratio shows clear peaks at different photon energies, which point to certain defects in the diamond: The most prominent lines appear at \SI{2.2}{\eV} and \SI{2.54}{\eV}, which we attribute to substitutional nitrogen (N$_s^0$) and divacancies respectively.
\textbf{D} Conversion voltage as a function of the excitation photon energy. The conversion is most efficient at \SI{2.4}{\eV}, which we attribute to a trade off between efficient excitation to $^2$A$_{\rm{2u}}$ and efficient hole generation via divacancies, which have an absorption band at $\widehat{=}$\SI{2.54}{\eV} \cite{Pu2001_Divacancy_488nm}. 
} \end{figure*}

Figure \ref{fig:wavelength-dependence-switching}A shows the integrated ZPL intensity of the SiV$^-$ emission as a function of the bias voltage measured with a spectrometer and with the green excitation laser focused close to the gold-diamond interface. It decreases for negative biases with a behavior that can be well described by a sigmoid fit with constant offset. We define the conversion voltage as the center of the sigmoid function. This PL decrease is indicative of a progressive transition away from the bright SiV$^{-}$ to either the dark SiV$^{2-}$ or bright SiV$^{0}$, which emits at \SI{1.31}{\eV} \cite{Rose2018_SiV0_250msCoherence} and hence does not contribute to PL$_\mathrm{SiV^-}$, the integrated zero-phonon line of the SiV$^-$. 
Due to the absence of the SiV$^{0}$ signature at \SI{1.31}{\eV}, we conclude that SiV$^{-}$ is converted to SiV$^{2-}$ at the negative electrode, while SiV$^{-}$ is preserved at the positive contact. 
We note that we do not observe any indication of linear Stark shifts as shown in the inset of figure \ref{fig:wavelength-dependence-switching}A which is expected for inversion symmetric SiV \cite{Sipahigil2014_Indistinguishability, GaliMaze, Aghaeimeibodi2021_tinVacancyStarkTuning}.

To characterize the low-temperature excitation spectrum, we performed spectrally-resolved PL-excitation measurements by tuning the excitation photon energy from $1.77$ to \SI{2.75}{\eV} and detecting the SiV$^-$ emission. Figure \ref{fig:wavelength-dependence-switching}B presents PL excitation spectra recorded with a bias of \SI{20}{\volt} with the laser spot positioned at the midpoint between two electrodes and using ultrafast pulses (pulse energy of \SI{125}{\pico\joule}, \SI{80}{\mega\hertz} repetition rate and $80$ to \SI{200}{fs} pulse duration). We find that the SiV$^-$ centers are excited throughout this spectral window, but most efficiently at \SI{2.27}{\eV}, where we observe a peak. Given the discrete set of ground state levels, the measured broad excitation band indicates that the excitation process involves multiple excitation pathways as well as phonon mediated processes. At \SI{2.27}{\eV}, we observe a pronounced resonance with a linewidth of \SI{43(3)}{\milli\eV} comparable to the bandwidth of the excition sources ($\geq\SI{20}{\milli\eV}$, measured in spectrometer). We attribute this peak to resonant transitions from the deeper SiV$^-$ ground state level $a_{2u}$ to the excited state level $e_g$ \cite{Haußler2017_a2u_level, GaliMaze}.

Figure \ref{fig:wavelength-dependence-switching}C shows the excitation photon energy-dependent ideality factor ratio $n_1/n_2$ defined in equation \ref{eq:schottkybarrier}, where subscripts $1,2$ denote the electrodes at the Schottky contact, extracted from the photocurrent data presented in fig.\ref{fig:SiV-PL_sample_Map-cut}. Unlike the rest of the experiments, those measurements were carried out with a large laser beam waist ($1/e^2$ waist of \SI{28}{\micro\meter}).
Generally, the ideality factor characterizes the junction quality providing a measure of how closely the diode follows the ideal diode equation. Recently, it has been shown that the ideality factor contains useful information on the defect levels at the active region of the Schottky barrier \cite{shin2022understanding}.
Thereby, we identify several resonances in the ideality factor arising from resonant interactions with defects in the diamond lattice, such as N$_s^0$ centers (substitutional nitrogen) and divacancies \cite{Rosa1999_substNitrogen,Dyer1965_substNitrogen,Pu2001_Divacancy_488nm}.
Substitutional nitrogen (divacancy) centers act as electron donors (acceptors) in pristine diamond. 
The observed peak at \SI{2.54}{\eV} suggests that a significant number of divacancies is present in our diamond and therefore aids the conversion mechanism for the SiV$^{2-}$ to SiV$^-$ conversion.  We propose that they act as electron acceptors under optical excitation, thereby providing free holes in the vicinity of the SiV$^{2-}$ centers, while the quasi-resonant transition from the lower ground state $a_{2u}$ level assists the SiV$^{-}$ to SiV$^2-$ conversion. As a consequence of the combination of the two aforementioned efficiencies, the overall conversion efficiency peaks at \SI{2.4}{eV}, as illustrated in figure \ref{fig:wavelength-dependence-switching}D which quantifies the conversion voltage.

We now determine the conversion voltage from the bias-dependent photoluminescence for different excitation wavelengths at constant $P_{\mathrm{ex}}$ = \SI{10}{\milli\watt} (see figure S5 in the Supplemental for all voltage dependent PL spectra). In the range from $2.70$ to \SI{2.25}{\eV}, we observe charge state conversion, with varying conversion voltages determined from sigmoid fits. Figure \ref{fig:wavelength-dependence-switching}D presents the results of this analysis and shows that the conversion voltage is the lowest for excitation with \SI{2.4}{eV} light. Both positive and negative bias voltages increase SiV$^-$ PL intensity for photon energies above \SI{2.7}{\eV}, indicating that these energies partially convert the charge state to the dark state under zero electrical bias (see figure S5 in the Supplemental). For excitation below \SI{2.25}{\eV}, we can not observe the conversion up to \SI{200}{\volt}, indicating that a process in the conversion is inefficient for such excitation wavelengths (also figure S5). The strong wavelength dependence furthermore clearly confirms that the conversion mechanism is not primarily driven by Fermi level shifting but instead by optically created charge carriers and excitation of the SiV$^-$.

\subsection*{Time-resolved interconversion between SiV$^-$ and SiV$^{2-}$}

\begin{figure*}
\centering
\includegraphics[width=\textwidth]{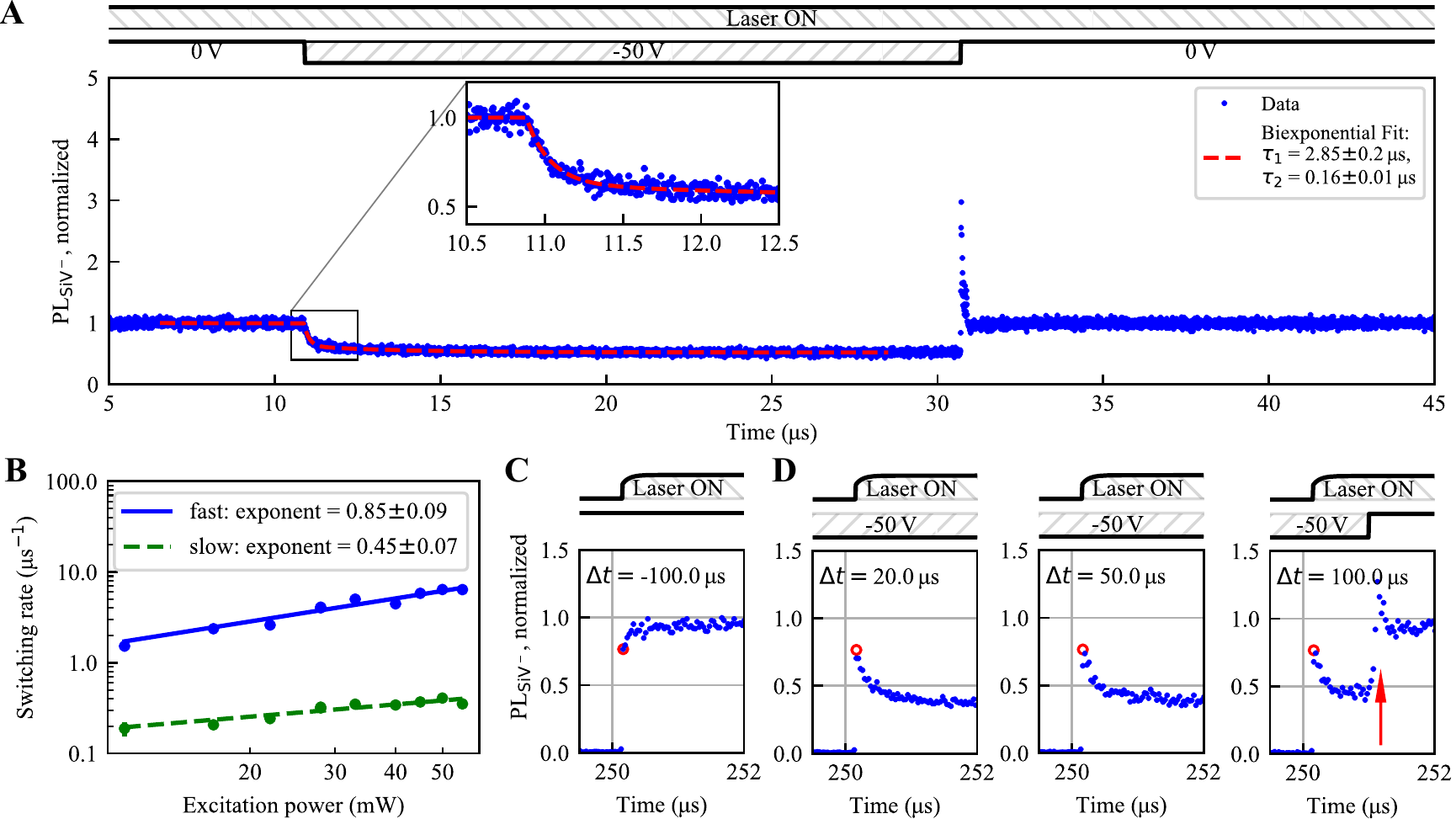}
\caption{\label{fig:Time-resolved-switching_turnOff}
\textbf{Time-resolved measurements of the SiV$^-$ $\leftrightarrow$ SiV$^{2-}$ interconversion.}
\textbf{A} While we illuminate continuously with \SI{2.40}{\eV} photon energy, we apply \SI{20}{\micro\second}-long voltage pulses and  correlate them temporally with the SiV$^-$ PL measured with a single photon detector. As the bias voltage is switched from \SI{0}{\volt} to \SI{-50}{\volt}, the SiV$^-$ PL decreases as SiV$^-$ convert to SiV$^{2-}$. A biexponential fit (red dashed line) describes the decay very well (see the legend for the time constants). 
\textbf{B} The conversion rates from the biexponential fit scale with the excitation power $P_{\mathrm{ex}}$. The fast decay reaches \SI{}{MHz} conversion rates for strong illumination.
\textbf{C} Digitally modulated laser and delayed  relative to the voltage pulse. The laser pulse starts after the voltage pulse ended. The red circle marks the first data point during the laser pulse and serves as a reference for the following measurements. 
\textbf{D} Laser pulse starting during the voltage pulse. We still observe the biexponential decay, which starts exactly from the baseline PL marked with the red circle. We find constant onset PL for relative pulse delays of \SI{20}{\micro\second}, \SI{50}{\micro\second} and \SI{100}{\micro\second}, which indicates that purely electrical SiV charge state conversion is insignificant on a microsecond time scale. The arrow marks what we call PL overshoot.
} \end{figure*}

We continue to study the temporal dynamics of the charge state interconversion between SiV$^-$ and SiV$^{2-}$. For this, as depicted schematically in figure \ref{fig:Time-resolved-switching_turnOff}A, we apply a pulse pattern of \SI{-50}{\volt} voltage pulses from a DC level of \SI{0}{\volt} in order to switch the system between regimes where the SiV$^-$ emission intensity is high (\SI{0}{\volt}) and low (\SI{-50}{\volt}).  A CW laser at \SI{2.40}{\eV} is continuously focused close to the negative electrode. The PL emission is correlated temporally with the voltage pulse by measuring with a single photon avalanche diode using a boxcar method. We find that the PL intensity reduces upon switching the voltage from \SI{0}{\volt}$\rightarrow$\SI{-50}{\volt} and a clear transient is observed as SiV$^-$ centers are converted to SiV$^{2-}$. We fit this transient using a biexponential decay to a fraction of the unbiased PL without electrical bias. This results in a $\sim$50\% conversion contrast, consistent with the expected contrast visible in figure \ref{fig:SiV-PL_sample_Map-cut}B for \SI{-50}{\volt} bias.
In figure \ref{fig:Time-resolved-switching_turnOff}B the "turn-off" conversion rates (defined as the inverse of the biexponential decay constants) are plotted versus $P_{\mathrm{ex}}$. We find that both rates scale according to a power law with exponents of $0.85 \pm 0.08$ for the fast decay and $0.44 \pm 0.06$ for the slow decay, respectively. For the highest $P_{\mathrm{ex}}$ we reach \SI{}{MHz} conversion rates for the fast turn-off, which is the dominant effect converting the majority of the bright SiV$^-$ to the dark SiV$^{2-}$ state, see figure S7 for the relative amplitudes of the biexponential decay. The slower process could come from centers undergoing multiple charge state changes due to continuous excitation and holes from other SiV$^-$ to SiV$^{2-}$ transformations or from the spatially varying power of the Gaussian beam profile, but such processes are secondary.

To study purely electrical charge state interconversion, we additionally gate the CW Laser and vary the relative delay to the voltage pulses. Figure \ref{fig:Time-resolved-switching_turnOff}C shows a baseline measurement with zero electrical bias where we see that luminescence ramps up to a steady state value within less than \SI{1}{\micro\second}. Now we let a voltage pulse (\SI{-50}{\volt}) precede and overlap with the laser pulse, with varying overlap. A purely electrical charge state conversion should decrease the transient luminescence amplitude, but this is not observed. In contrast, the initial amplitude seems to be completely unchanged, which is presented in figure \ref{fig:Time-resolved-switching_turnOff}D for varying overlaps (\SI{20}{\micro\second}, \SI{50}{\micro\second} and \SI{100}{\micro\second}). Purely electrical charge state conversion is therefore negligible. 

\begin{figure*}
\centering
\includegraphics[width=\textwidth]{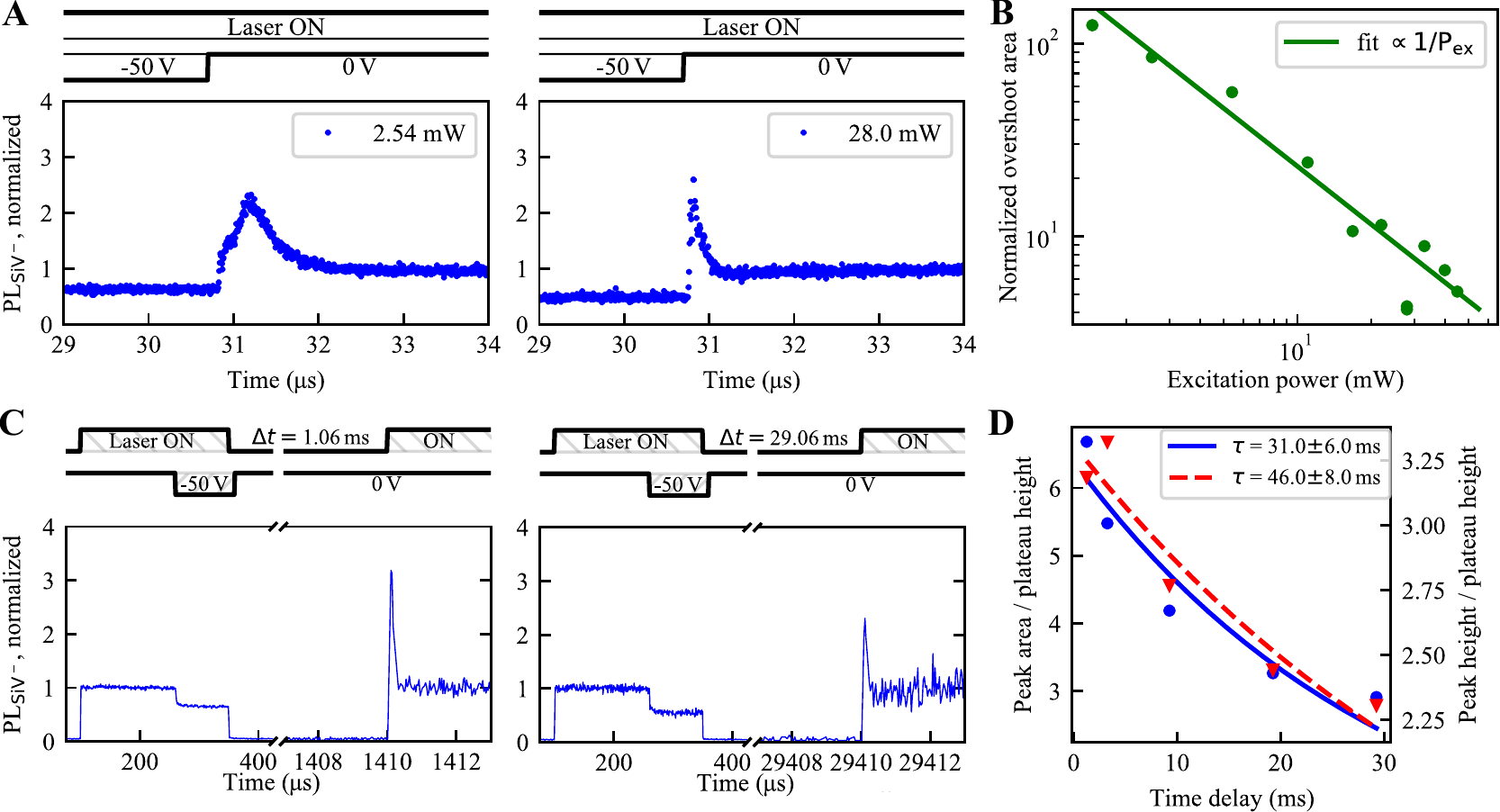}
\caption{\label{fig:timeResolved_turnON_power} 
\textbf{Power dependence of the PL overshoot during the conversion from SiV$^{2-}$ to SiV$^-$ and dark relaxation of SiV$^{2-}$ to SiV$^-$.}
\textbf{A} The sample is excited with a CW \SI{2.4}{\eV} laser with a power $P_{\mathrm{ex}}$ of \SI{2.54}{\milli\watt} or \SI{28}{\milli\watt}. A \SI{-50}{V} bias pulse first converts a fraction of SiV$^-$ to SiV$^{2-}$. After the bias is switched off, the SiV$^-$ PL rapidly increases, exhibiting an overshoot. As discussed in the main text this arises from hole capture of SiV$^{2-}$ resulting in an SiV$^-$ excited state. The PL overshoot becomes temporally narrower with increasing $P_{\mathrm{ex}}$ since the optically created hole flux increases, thereby, more rapidly driving the conversion. 
\textbf{B} The area of the PL overshoot is divided by the SiV$^-$ PL count rate at zero bias to normalize it. The result scales inversely with $P_{\mathrm{ex}}$. The SiV$^-$ PL intensity is linearly proportional to $P_{\mathrm{ex}}$ indicating that the number of photons in the PL overshoot is independent of $P_{\mathrm{ex}}$. This is the expected behavior as we expect a single photon from each SiV$^{2-}$ to SiV$^-$ conversion.
\textbf{C} After converting to the SiV$^{2-}$ state by applying \SI{-50}{\volt} and a \SI{2.4}{\eV} laser, the laser is turned off. When the voltage is switched off subsequently, we do not observe a PL overshoot. The delay between voltage turn-off and laser turn-on in the plots is \SI{1.06}{\milli\second} and \SI{29.06}{\milli\second}. We see that the intensity of the PL overshoot slowly decreases with the delay, which indicates a relaxation of the non-equilibrium SiV$^{2-}$ population to SiV$^-$. 
\textbf{D} The PL overshoot intensity decreases with a time constant of \SI{36.4\pm6.7}{\milli\second} in the dark (weighted average of the two exponential decay constants). This can be interpreted as the lifetime of the SiV$^{2-}$ charge state in the dark.
} \end{figure*}

The SiV$^-$ $\rightarrow$ SiV$^{2-}$ conversion at the \SI{-50}{\volt}$\rightarrow$\SI{0}{\volt} transition results in a PL overshoot. The rightmost panel of figure \ref{fig:Time-resolved-switching_turnOff}D shows this overshoot and that the PL signal rapidly increases to the saturated level over a timescale faster than the temporal resolution of our measurement ($\sim\SI{100}{\nano\second}$). As presented in figure \ref{fig:timeResolved_turnON_power}A, the overshoot becomes temporally shorter for higher excitation levels.
On the rightmost panel in figure \ref{fig:Time-resolved-switching_turnOff}D, the conversion from SiV$^{2-}$ to SiV$^-$ can be observed at the transition from \SI{-50}{\volt} to \SI{0}{\volt}, accompanied by an increase of SiV$^-$ PL. From now on we refer to this process as the "turn-on" and observe that it happens significantly faster than the turn-off. Furthermore, we see a short overshoot of SiV$^-$ PL emission at the turn-on marked by the arrows on the figure, which we term PL overshoot from now on. As presented in figure \ref{fig:timeResolved_turnON_power}A, the overshoot becomes temporally shorter for higher $P_{\mathrm{ex}}$. Figure \ref{fig:timeResolved_turnON_power}B presents the area of this PL overshoot as a function of $P_{\mathrm{ex}}$. The integrated PL overshoot intensity is found to be inversely proportional to $P_{\mathrm{ex}}$ if normalized to the unbiased PL count rate. The latter is linearly proportional to $P_{\mathrm{ex}}$, which means that the absolute number of photons in the PL overshoot is independent of $P_{\mathrm{ex}}$. This indicates that the PL overshoot intensity depends only on the number of non-equilibrium SiV$^{2-}$ that are re-converted to SiV$^-$. 

To investigate the role of the excitation laser in the PL overshoot, we performed experiments in which we turn the laser off before switching the voltage back from \SI{-50}{\volt} to \SI{0}{\volt}.  Typical results of these investigations are presented in figure \ref{fig:timeResolved_turnON_power}C. We do not observe the PL overshoot during this voltage switch-off. Instead, after a variable delay, we turn the laser back on and again observe the PL overshoot. These observations show that the PL overshoot is optically triggered, which also means that laser illumination is \textit{required} for the SiV$^{2-}$ to SiV$^-$ conversion.
Nevertheless, the voltage switch-off perturbs the relative equilibrium populations of SiV$^-$ and SiV$^{2-}$ and it is expected that some SiV$^{2-}$ slowly relax back to SiV$^-$ after applying the voltage pulse. By varying the delay between voltage turn-off and laser turn-on, we measured the time scale over which this process occurs, namely the lifetime of the non-equilibrium SiV$^{2-}$ charge state under dark conditions. The results of these experiments are presented in figure \ref{fig:timeResolved_turnON_power}D. In this figure we normalized the PL overshoot intensity to the equilibrium SiV$^-$ intensity and fitted the intensity versus delay with a mono-exponential decay, from which we determined the lifetime of the SiV$^{2-}$ charge state to be \SI{36\pm7}{\milli\second} in the dark. 
Comparing this to the $\ll$\SI{500}{\nano\second}-long optically driven conversion to SiV$^-$, we find that optical excitation accelerates this process by a factor of $\sim10^5$ clearly underscoring the importance of optical pumping in the conversion of dark SiV$^{2-}$ to the bright SiV$^-$ state. 

\section*{Discussion}

\subsection*{Charge state conversion mechanisms}

\begin{figure*}
\centering
\includegraphics[width=0.7\textwidth]{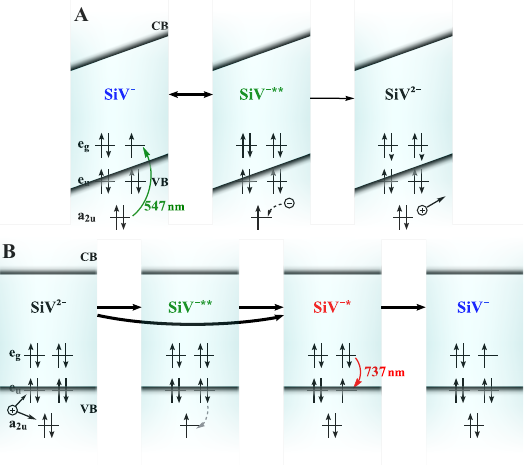}
\caption{\label{fig:chargeCycleSwitching} 
\textbf{Proposed mechanism of the charge state interconversion between SiV$^{-}$ and SiV$^{2-}$ and luminescence by hole-capture.}
\textbf{A} Conversion from SiV$^{-}$ to SiV$^{2-}$. Starting from the SiV$^-$ ground state, a photon excites an electron from the lower lying SiV$^-$ level $a_{2u}$ to $e_g$, which results in a higher excited state SiV$^{-**}$ ($^2$A$_{\rm{2u}}$). A VB electron tunnels into the $a_{2u}$ level, resulting in SiV$^{2-}$ and a free hole. As the tunnel probability is constant, we expect an overall linear dependence on $P_{\mathrm{ex}}$, which we indeed observed for the fast dominant process in figure \ref{fig:Time-resolved-switching_turnOff}B.
\textbf{B} Conversion from SiV$^{2-}$ to SiV$^{-}$ and PL overshoot. A free hole (e.g. from the positive electrode interface or a divacancy) is captured by $e_u$ or $a_{2u}$ which results in an excited SiV$^-$ state ($^2$E$_{\rm{u}}$ or $^2$A$_{\rm{2u}}$ respectively). $^2$A$_{\rm{2u}}$ will relax to $^2$E$_{\rm{u}}$, which emits a photon at \SI{1.68}{\eV}, causing the PL overshoot. This model also explains the behavior in figure \ref{fig:SiV-PL_sample_Map-cut}D where we saw that the SiV$^-$ intensity is the highest if the current and hence the hole density is the highest. Note that the energy separation between the levels are not to scale and that the relative position of the $e_u$, $e_g$ and $a_{2u}$ to the diamond bands depend on excitation and charge state, which is not shown for the sake of simplicity.}
\end{figure*}

To explain the SiV charge state interconversion, we propose the mechanism depicted schematically in figure \ref{fig:chargeCycleSwitching}. The involved levels of SiV include, in energetically ascending order, $a_{2u}$ deep in the valence band followed by $e_u$ which is near the band gap according to DFT calculations and $e_g$, which is inside the band gap \cite{GaliMaze}. In the SiV$^-$ ground state $^2$E$_{\rm{g}}$, there is a single hole in $e_g$, while the other two levels are completely filled \cite{GaliMaze}. Figure \ref{fig:chargeCycleSwitching}A shows the SiV$^-$ to SiV$^{2-}$ conversion, where a \SI{2.27}{\eV} photon resonantly excites an electron from the $a_{2u}$ to the $e_g$ level, leaving the SiV$^-$ in its SiV$^{-**}$ (or $^2$A$_{\rm{2u}}$) excited state. From there, the system can either relax back to the ground state $^2$E$_{\rm{g}}$ which causes SiV$^-$ PL or an electron from the VB can repopulate the hole in $a_{2u}$. The latter mechanism directly converts the SiV$^{-**}$ to SiV$^{2-}$, leaving a hole in the diamond VB. When we apply a bias voltage, the VB hole is attracted to the negatively biased electrode, suppressing capture by the SiV and thus stabilizing the vacancy in the dark SiV$^{2-}$ charge state. However, from our discussion of the PC data above, and as discussed in the context of figure \ref{fig:SiV-PL_sample_Map-cut}D and \ref{fig:SiV-PL_sample_Map-cut}E, we know that we optically create holes near the positive electrode and in between the electrodes, which can also drive the conversion back to SiV$^{-}$. As a consequence, we expect a higher fraction of SiV$^-$ and therefore more SiV$^-$ PL near the positive electrode, which is indeed what we observed in figure \ref{fig:SiV-PL_sample_Map-cut}D. Similarly to the PC, the PL also peaks at the positive electrode, but the PL peak is spatially shifted away from the electrode. This can be explained by partial reflection of laser light and PL (both of which have a Gaussian beam profile) from the metal electrode. 

Similar as in the purely optically driven charge state conversion cycle of tin vacancies \cite{TinVacancyCycle_Becher2022}, also SiV$^{2-}$ could directly recapture the created VB hole and restore the SiV$^-$ charge state. However, this is no longer possible when the hole is attracted by the negative electrode and thus separated from SiV$^{-}$ immediately. 
For future applications and experiments with SiV$^-$, we therefore propose an electrically assisted charge state conversion and stabilization instead of the standardly used purely optical method.

\subsection*{SiV$^-$ luminescence overshoot by hole capture}

We continue to discuss the underlying physical mechanism driving the SiV$^{2-}$ to SiV$^{-}$ conversion, depicted in figure \ref{fig:chargeCycleSwitching}B. To convert the SiV back to the SiV$^-$ state, a VB hole needs to be recaptured by an SiV$^{2-}$ center.
The hole could either be captured by $e_u$, which lies close to the VB edge \cite{GaliMaze}, or it could be captured by $a_{2u}$, which requires additional energy of the hole since the level lies deeper in the VB. From the implantation dose, we expect to have $\approx100$ SiV per \SI{}{\micro\meter^2}, which means that the average distance between two SiV is $\approx$\SI{100}{\nano\meter}. For a bias voltage of \SI{50}{V} and an electrode distance of \SI{8}{\micro\meter}, a hole therefore gains $\approx\SI{100}{\nano\meter}/\SI{8}{\micro\meter}\cdot\SI{50}{\volt}=\SI{0.6}{\eV}$.  This energy is comparable to the energetic separation between $a_{2u}$ and $e_u$, which is $\approx\SI{2.27}{\eV}-\SI{1.68}{\eV}=\SI{0.6}{\eV}$. 
Of course the situation is further complicated by scattering with other defects and the limited capture probability of a hole in an SiV. Regardless of whether the hole is captured into $a_{2u}$ or $e_u$, this automatically results in an excited state of the SiV$^-$ and hence PL at \SI{1.68}{\eV} in full accord with our experimental observations. 
As shown in figure \ref{fig:chargeCycleSwitching}B, we expect SiV$^{-**}$ to relax to the SiV$^{-}$ ground state in two steps via SiV$^{-*}$, so that both capturing processes lead to the emission of a \SI{1.68}{\eV} photon.

This is quite different for the recovery of SiV$^-$ from SiV$^{0}$, which requires the capture of an electron and leaves SiV$^-$ directly in the ground state without producing additional PL. We therefore take the observed PL overshoots that emerge upon charge state conversion as a strong indication that we indeed cycle between SiV$^{-}$ and SiV$^{2-}$ instead of between SiV$^{-}$ and SiV$^{0}$.
Thereby, the PL overshoot stems from optically created holes and should therefore change with the laser illumination. The total PL overshoot intensity depends primarily on the number of non-equilibrium SiV$^{2-}$ that we created opto-electronically. With a higher flux of VB holes, we expect this population to reconvert faster. We indeed find that the PL overshoot becomes temporally narrower for higher $P_{\mathrm{ex}}$, see figure \ref{fig:timeResolved_turnON_power}A. Also, since the total intensity of the PL overshoot should be roughly independent of $P_{\mathrm{ex}}$, we should observe that the overshoot intensity becomes larger relative to the steady-state PL (which is linearly proportional to $P_{\mathrm{ex}}$), which is also what we see in our data in figure \ref{fig:timeResolved_turnON_power}B.

This charge state conversion process is much more efficient at the positive electrode that promotes the generation of free holes in the VB and now we discuss how this can be used to stabilize the SiV$^-$ charge state. Figure \ref{fig:SiV-PL_sample_Map-cut}D shows that SiV$^-$ PL increases by \SI{20}{\%} at the positive electrode under $\SI{60}{V}$ DC bias. This is most likely caused by SiV$^{2-}$ that are converted to SiV$^-$ and produce additional PL. 
A potential protocol for maximizing the SiV$^-$ population would therefore consist of a green laser pulse and positive electrical bias that can be continuously applied.

\subsection*{Final discussion}

Our experimental results provide a better understanding of the photophysics of SiV color centers charge exchange with the diamond host and provide the basis for changing and stabilizing their charge states using local electric fields. Due to the similarity in the energy level structure and position in the band gap \cite{Thiering2018_MagnetoOpticalSpectra_G4V}, the findings are most certainly also valuable for understanding other types of G4V centers.
Furthermore, they aid the development of an opto-electronic spin-readout technique, following the well-established electrical readout of NV centers in diamond, which is also based on spin-dependent charge state conversion processes \cite{NVreadoutMartinBrandt, Bourgeois2015NatComm}. These techniques are especially favorable for building compact electronic devices and for overcoming optical spin readout, which is inefficient due to the high refractive index of diamond \cite{ShieldsLukin2015PRL, Bourgeois2015NatComm}.

Last, we discuss the limitations of this study and our view on the next steps towards fully charge-stabilized SiV. 
For charge state stabilization protocols, the current density needs to be small in the vicinity of the SiV spin qubits, as the electric fields of moving charges can cause decoherence. We measured very low dark currents smaller than \SI{1}{\nano\ampere} for the whole interdigitated electrodes and hence, in combination with the inversion symmetry of the SiV, do not expect them to be an issue. However, this should be experimentally verified and strong local photocurrents could have an influence on coherence, such that we recommend a pulsed stabilization scheme without illumination during the qubit operation.
In such a protocol, the charge state has to be conserved for longer than the relevant $T_2$ time of the system. We measured the charge state relaxation time to be \SI{36.4\pm6.7}{\milli\second} at \SI{8}{\kelvin}, which is longer than the \SI{10}{\milli\second} electron spin coherence time $T_2$ with dynamical decoupling. However, nuclear spin coherence times can reach the timescale of seconds, causing stronger requirements. We expect the charge state relaxation time to increase towards millikelvin temperatures, which should be verified. 
Furthermore, the Schottky contacts need to be integrated into cavities to get efficient spin-photon interfaces, for example by using an evanescently coupled cavity and waveguide system with in-plane surface electrodes on the diamond \cite{Craiciu2021_evanescentCavity,Ourari2023_evanescentCavity}. 

\section*{Materials and methods}

\subsection*{Sample preparation}\label{sec:sample}

To create silicon vacancy centers, we implanted high-purity, single-crystal diamond produced by Element Six. The diamond was grown using Chemical-Vapor-Deposition (CVD) with a size of \SI{3}{\milli\meter} $\times$ \SI{3}{\milli\meter} $\times$  \SI{0.5}{\milli\meter}. This type of diamond typically contains nitrogen and boron impurity levels $<\SI{5}{ppb}$ and $<\SI{0.5}{ppb}$, respectively \cite{E6_handbook}. 

$^{28}$Si was implanted with \SI{132}{\kilo\eV} by CuttingEdge Ions (expected implantation depth of \SI{100}{} to \SI{150}{\nano\meter}) with a dose of \SI{3e11}{ions/\centi\meter^{-2}}. After implantation, we annealed in several steps ($\SI{4}{\hour}$ ramp to \SI{400}{^\circ C}, $\SI{8}{\hour}$ at \SI{400}{^\circ C}, $\SI{12}{\hour}$ ramp up to \SI{800}{^\circ C}, $\SI{8}{\hour}$ at \SI{800}{^\circ C}, $\SI{12}{\hour}$ ramp to \SI{1100}{^\circ C}, $\SI{10}{\hour}$ at \SI{1100}{^\circ C}) under high vacuum ($<\SI{1e-7}{mbar}$) \cite{Evans2016_annealingProcedure}. Throughout the process, all temperature increments were executed gradually to preserve a high vacuum within the annealing chamber, essential for avoiding surface graphitization of the diamond \cite{ruf2019optically}.

We cleaned the sample in aqua regia ($\SI{1}{\hour}$ at \SI{90}{\celsius}) and concentrated sulphuric acid (1h at \SI{225}{\celsius}), followed by exposure to oxygen plasma which produces oxygen termination. Electrodes of \SI{8}{\micro\meter} distance were created by optical lithography and evaporation of Ti(\SI{5}{\nano\meter})/Au(\SI{80}{\nano\meter}).
The interdigitated electrodes are each connected to a \SI{200}{\micro\meter}x\SI{200}{\micro\meter} bond pad. The sample is glued to a copper plate and wire-bonded to a printed circuit board chip carrier (see figure S1 in the Supplemental). Electrical DC measurements were performed with a Keithley 2400 source-measure unit. Electrical pulses were created with a HP 218B pulse generator with a rise time of \SI{10}{\nano\second}. The sample was cooled to \SI{8}{\kelvin} using a helium flow cryostat.

\subsection*{Optical and electrical measurements}

The optical measurements in this work are performed with a custom-built confocal microscope photoluminescence setup. The excitation Laser (Toptica iBeam smart, \SI{517}{\nano\meter} with digital intensity modulation function) is focused with a 100x magnification, $0.5$ NA, broadband objective. Wavelength-scans were performed with a Spectra physics tunable HF100 optical parametric oscillator (OPO) system which emits ultrashort pulses with a duration of \SI{80}{\femto\second} to \SI{200}{\femto\second} and a repition rate of \SI{80}{\mega\hertz}. 

Pulse patterns were created with a Pulse Streamer 8/2 from Swabian instruments. We used a spectrometer with \SI{500}{\milli\meter} focal length for continuous measurements or a fiber coupled single photon avalanche photodiode (SPAD) for detection in time-resolved experiments. The excitation laser was separated from the signal by using a spectral $732$ to \SI{742}{\nano\meter} bandpass filter combined with \SI{532}{\nano\meter}, \SI{600}{\nano\meter} or \SI{700}{\nano\meter} longpass filters, depending on the excitation wavelength. For time-resolved measurements we also used a tunable longpass and a tunable shortpass filter to filter the SiV$^-$ emission tightly around the zero-phonon line. Temporal correlations were taken with the time-tagger QuTag from qutools.


\bibliography{scibib}
\bibliographystyle{Science}

\section*{Acknowledgments}

This work was supported by the German Science Foundation DFG via the clusters Munich Center for Quantum Science and Technology MCQST and e-conversion (EXC 2111 and EXS 2089, resp.) as well as via the instrumentation projects PQET (INST 95/1654-1) and MQCL (INST 95/1720-1), by the German Federal Ministry of Education and Research BMBF via the projects SPINNING (13N16214), QuaDiQua (16K1S0948) and epiNV (13N15702), by the Bavarian State Ministry of Science and Arts via the project EQAP and by Bayerisches Staatsministerium für Wissenschaft und Kunst through project IQSense via the Munich Quantum Valley MQV.
V. Villafañe gratefully acknowledges the Alexander v. Humboldt foundation and MCQST for financial support in the framework of their fellowship programs.

\section*{Supplementary materials}
Figs. S1 to S8\\


\clearpage


\end{document}